\documentclass{aa}  

\usepackage{graphicx}
\usepackage{txfonts}
\usepackage{booktabs}
\usepackage[para]{threeparttable}
\usepackage{xcolor}
\begin{document}

   \title{LoTSS Jellyfish Galaxies}

   \subtitle{III. The first identification of jellyfish galaxies in the Perseus cluster}

   \author{I.D. Roberts
          \inst{1}
          \and
          R.J. van Weeren\inst{1}
          \and
          R. Timmerman\inst{1}
          \and
          A. Botteon\inst{1}
          \and
          M. Gendron-Marsolais\inst{2}
          \and
          A. Ignesti\inst{3}
          \and
          H.J.A. Rottgering\inst{1}
          }

   \institute{Leiden Observatory, Leiden University, PO Box 9513, 2300 RA
Leiden, The Netherlands\\
        \email{iroberts@strw.leidenuniv.nl}
        \and
            European Southern Observatory, Alonso de Córdova 3107, Vitacura, Casilla 19001, Santiago de Chile, Chile
        \and
            INAF- Osservatorio astronomico di Padova, Vicolo Osservatorio 5, IT-35122 Padova, Italy\\
}

\abstract{In this paper we report the first identification of jellyfish galaxies in the Perseus cluster (Abell 426).  We identified four jellyfish galaxies (LEDA 2191078, MCG +07-07-070, UGC 2654, UGC 2665) within the central $2^\circ \times 2^\circ$ ($2.6\,\mathrm{Mpc} \times 2.6\,\mathrm{Mpc}$) of Perseus based on the presence of one-sided radio continuum tails that were detected at $144\,\mathrm{MHz}$ by the LOw Frequency ARray (LOFAR). The observed radio tails, as well as the orientation of morphological features in the rest-frame optical, are consistent with these four galaxies being impacted by ram pressure stripping as they orbit through the Perseus intracluster medium.  \textcolor{black}{By combining the LOFAR imaging at 144 MHz with 344 MHz imaging from the Karl G. Jansky Very Large Array, we derived spectral indices for the disks and the stripped tails of these jellyfish galaxies.  We show that the spectral indices over the galaxy disks are quite flat, while the indices of the stripped tails are substantially steeper.  We also identified a number of compact $\mathrm{H\alpha + [N\textsc{ii}]}$ sources with narrowband imaging from the Isaac Newton Telescope.  These sources are brighter along the leading side of the galaxy (i.e.,\ opposite to the direction of the stripped tail), which is consistent with ram pressure induced star formation.  Lastly, consistent with previous works in other clusters, we find that these jellyfish galaxies show enhanced radio luminosities for their observed star formation rates.} Given the small distance to the Perseus cluster ($D \sim 70\,\mathrm{Mpc}$, $1'' \simeq 340\,\mathrm{pc}$), these galaxies are excellent candidates for multiwavelength follow-up observations to probe the impact of ram pressure stripping on galaxy star formation at subkiloparsec scales. \\}

\date{Received September 15, 1996; accepted March 16, 1997}

\maketitle

%

\section{Introduction} \label{sec:intro}

It is now firmly established that the extreme environments of galaxy clusters strongly influence the properties of galaxies within. This is seen most easily via the overabundance of red, passive galaxies in clusters relative to low-mass groups or the field \citep[e.g.,][]{dressler1980,haines2006,blanton2009,wetzel2012,haines2015,brown2017,roberts2019}. The presence of this ``environmental quenching'' has been cemented by large redshift surveys of the local Universe, with focus now shifting toward understanding the physical drivers of these observed trends.  Many physical mechanisms capable of quenching galaxies in clusters have been proposed. These processes can be divided between gravitational interactions like mergers \citep[e.g.,][]{mihos1994a,mihos1994b}, tidal interactions \citep[e.g.,][]{mayer2006,chung2007}, or impulsive galaxy encounters \citep[`harassment', e.g.,][]{moore1996}, and hydrodynamic interactions like ram pressure stripping \citep[RPS, e.g.,][]{gunn1972,quilis2000,poggianti2017}, viscous stripping \citep[e.g.,][]{nulsen1982,quilis2000}, or starvation \citep[e.g.,][]{larson1980,peng2015}.
\par
Increasingly, RPS has been suggested to play an important role in this environmental quenching \citep[e.g.,][]{gavazzi2001,boselli2006,chung2007,yagi2010,jachym2014,muzzin2014,kenney2015,boselli2018,brown2017,poggianti2017,maier2019,roberts2019,ciocan2020}.  Ram pressure can influence galaxy star formation both by removing cold gas (leading to rapid quenching) and by driving increased densities in the interstellar medium (ISM), potentially leading to starburst behavior \citep[e.g.,][]{schulz2001,bekki2014,vulcani2018_sf,roberts2020,cramer2021}. The latter is predicted to occur on the ``leading-side'' of galaxies undergoing RPS, in other words, the side of the galaxy directly encountering the intracluster medium (ICM), opposite to the stripped tail \textcolor{black}{\citep[e.g.,][]{gavazzi2001,boselli2021}}. In order to place strong constraints on the effect of RPS on star formation within galaxies (i.e.,\ Is star formation quenched from the outside-in? Does ram pressure enhance star formation along the leading edge?), it is critical to observe galaxies experiencing RPS at high physical resolution (i.e.,\ subkiloparsec) and at multiple wavelengths since star formation is a small-scale, multiwavelength phenomenon.
\par
Nearby galaxy clusters such as Virgo ($D \sim 15\,\mathrm{Mpc}$), Fornax ($D \sim 20\,\mathrm{Mpc}$), Perseus ($D \sim 70\,\mathrm{Mpc}$), and Coma ($D \sim 100\,\mathrm{Mpc}$) are all excellent targets for this as typical $\sim$arcsecond scale observations will reach subkiloparsec scales.  The galaxy populations of Virgo, Fornax, and Coma have been studied in relative detail at multiple wavelengths (\textbf{Virgo:} e.g., \citealt{chung2009,davies2010,boselli2011,boselli2018,brown2021}; \textbf{Fornax:} e.g., \citealt{davies2013,serra2016,iodice2016,iodice2017,sarzi2018,zabel2019}; \textbf{Coma:} e.g., \citealt{gavazzi1987_hi,bravo-alfaro2000,hammer2010,smith2010,yagi2010,smith2017,gavazzi2018,chen2020,cramer2021}). The same is not true for the Perseus cluster. Due to its low Galactic latitude, observations of the Perseus galaxy population are far fewer\footnote{However, AGN emission and the central mini-halo in Perseus are extremely well-studied at X-ray and radio wavelengths \citep[e.g.,][]{boehringer1993,fabian2000,fabian2011,simionescu2011,urban2014,gendron-marsolais2017,gendron-marsolais2020}.} and existing studies have tended to focus on the cluster core, mostly limited to optical wavelengths \citep[e.g.,][]{kent1983,conselice2002,conselice2003_perseus,penny2009}. Recently some effort has been spent to rectify this by cataloging the member galaxies of Perseus out to the cluster outskirts \citep{penny2011,wittmann2019,meusinger2020}. Both the earlier works focused on the cluster core, and these recent studies extending to larger cluster-centric radius, show Perseus to be a cluster strongly dominated by passive, early-type galaxies -- though late-type galaxies do become more common toward the cluster exterior \citep{meusinger2020}. This suggests that environmental quenching is a strong driver of observed galaxy properties in Perseus. There is ample opportunity for detailed, high resolution studies of galaxy star formation, and its quenching, in the Perseus cluster moving forward.
\par
In this paper we present the first systematic search for galaxies undergoing RPS in the Perseus cluster. We use $144\,\mathrm{MHz}$ radio continuum observations of Perseus from the LOw Frequency ARray \citep[LOFAR,][]{vanhaarlem2013,shimwell2017,shimwell2019} to search for jellyfish galaxies with one-sided low-frequency radio tails extending asymmetrically about the optical galaxy center. These observations trace cosmic ray electrons (CREs) accelerated by supernovae, which can then be stripped (along with magnetic fields) out of the galaxy disk by RPS. These radio continuum tails trace magnetic fields extending well outside of the stellar disk for distances up to tens of kpc. Here we present the detection of four such jellyfish galaxies, likely undergoing strong RPS, within the central $2^\circ \times 2^\circ$ of the Perseus cluster (roughly out to $R_{500}$). In Sect.~\ref{sec:data_methods} we describe the data used in this paper as well as the methods for identifying LOFAR jellyfish galaxies. In Sect.~\ref{sec:individual_gal} we discuss each of the four jellyfish galaxies, with a focus on their observed morphologies at multiple wavelengths.  \textcolor{black}{In Sect.~\ref{sec:spec_index} we present measurements of spectral indices for these galaxies, both over the galaxy disks and the stripped tails. In Sect.~\ref{sec:Ha_srcs} we present an analysis of compact $\mathrm{H\alpha + [N\textsc{ii}]}$ sources in these jellyfish galaxies.  In Sect.~\ref{sec:radio_lum_sfr} we consider the position of these four jellyfish galaxies with respect to the radio continuum vs. star formation rate (SFR) relation.} Finally, in Sect.~\ref{sec:conclusion} we discuss and briefly summarize the results of this work.
\par
We assume a standard cosmology of $\Omega_M = 0.3$, $\Omega_\Lambda = 0.7$, and $H_0 = 70\,\mathrm{km\,s^{-1}\,Mpc^{-1}}$. We take the redshift of the Perseus cluster to be 0.0179 and all galaxy cluster-centric projected radii are measured relative to the position of the brightest cluster galaxy, NGC 1275.

\section{Data \& Methods} \label{sec:data_methods}

Below we describe both the multiwavelength imaging and the spectroscopic galaxy sample that are analyzed in this paper.  These data products cover a wide range in wavelength including the rest-frame optical, narrowband $\mathrm{H\alpha + [N\textsc{ii}]}$ imaging, and two low-frequency radio continuum bands.

\subsection{LOFAR 144 MHz Imaging} \label{sec:lofar_img}

We used the pointing P049+41 from the LOFAR Two-metre Sky Survey \citep[LoTSS,][]{shimwell2017,shimwell2019} centered on RA=49.247 and Dec=41.380 observed on Nov 3, 2016. These LoTSS observations cover the 120-168 MHz band in 195.3125 kHz wide sub-bands with 64 channels and 1 sec integration time. The 8 hrs observation was bracketed by two 10 min calibrator scans on 3C196. The international LOFAR stations also participated in this observation.
\par
The data was first processed with the prefactor \citep{vanweeren2016,williams2016,degasperin2019} using a high-resolution model of 3C196 provided by A. Offringa. This model was re-scaled to place it on the \citet{scaife2012} flux density scale. In the prefactor step, the XX-YY phase offsets and bandpasses are determined. Next, we used the LOFAR long baseline pipeline from \citet{morabito2021} to apply the prefactor calibration solutions to the target field P049+41. We also phased shifted the dataset toward the direction of 3C84, the radio AGN associated with NGC 1275.
\par
Using a four component starting model for 3C84 (single point source plus three Gaussians), we performed 20 cycles of self-calibration to obtain a subarcsecond resolution model of the source. For the selfcal calibration we used the standard procedure detailed in \citet{vanweeren2020}. When solving, baselines shorter than 40 k$\mathrm{\lambda}$ were excluded given that there is a large amount of unmodeled flux on the short baselines and no phase-up of the core stations was performed. This self-calibration converged toward a robust image of the sources showing similar structure as has been observed at higher frequencies. The high-resolution study of 3C84 will be described in Timmerman et al. (in prep).
\par
We proceeded with the subtraction of the high-resolution 3C84 model from the visibilities. This very bright source otherwise severely limits the dynamic range when imaging and calibrating the larger field of the cluster at a lower resolution. After removing 3C84, we phase shifted the dataset back to its original phase center and averaged it to 2 channels per sub-band and 8 seconds integration time. These are the default averaging settings for LoTSS processing at 6 arcsecond resolution. We then calibrated the data with the DDF-pipeline used for the LoTSS survey \citep{tasse2021}. This produces an image corrected for direction dependent effects in 45 facets. The region around 3C84 in this image still suffered from poor image quality due to dynamic range limitations from the more extended lobes of 3C84 and surrounding mini-halo. The complex morphology of this bright emission results in inaccuracies in the deconvolution algorithm used in the DDF-pipeline.
\par
To further improve the image quality in the cluster region we employed the extract produce (also described in \citealt{vanweeren2020}). This resulted in a dataset where only the sources in a 2.17 degree square box centered on 3C84 were retained. We also applied a correction for the beam model in the direction of 3C84. The 10 brightest sources in this 2.17 degree box were again extracted and self-calibrated using multiscale deconvolution in WSClean \citep{offringa2014,offringa2017} to deal with the complex extended emission around 3C84. From these calibration solutions in 10 directions, we created complex gain screens covering the square box region using radial basis function interpolation. We finally imaged these data with the Image Domain Gridding (IDG) algorithm \citep{vandertol2018,veenboer2019} applying the gain screens. Primary beam corrections where also determined via the IDG algorithm.
\par
The result of this processing is a high quality $144\,\mathrm{MHz}$ image covering the central $\sim\!2^\circ \times 2^\circ$ of the Perseus cluster at $6''$ resolution and $100\,\mathrm{\mu Jy\,beam^{-1}}$ rms.  In this paper we present extracted cutout images of four newly identified jellyfish galaxies in Perseus. A detailed analysis of this full image, including the diffuse central radio emission around NGC 1275 will be presented in a future work.

\subsection{VLA 344 MHz Imaging} \label{sec:vla_imaging}

\textcolor{black}{To constrain the spectral properties of the galaxies identified in this work, we also make use of Karl G. Jansky Very Large Array (VLA) P-band (230-470 MHz) observations of the Perseus cluster from \citet{gendron-marsolais2017}.  We point to \citet{gendron-marsolais2017} for a full description of the VLA observations and data reduction, and here we only provide a brief summary of the procedure.  The full dataset consists of 13hr in the P-band, including A-, B-, and D-configurations.  In this work we only use the 5hr B-configuration observation.  The dataset consists of 58 scans with one 10min observation on 3C48 and two 10min observations on 3C147 for calibration.  The remaining observations are made up of 5-10min scans on centered on NGC 1275.}
\par
\textcolor{black}{Data reduction was performed with CASA (Common Astronomy Software Applications, version 4.6), including a pipeline developed to address both the strong presence of radio frequency interference (RFI) at these low frequencies as well as the bright AGN emission from 3C84.  The steps of this procedure are outlined in detail in \citet{gendron-marsolais2017}. The final image is produced with a Briggs robust parameter of 0 and has a size of $\sim 5.12^\circ \times 5.12^\circ$ with a full-width at half-power of $\sim 2.4^\circ$.  The image has an elliptical beam with FWHM along the major and minor axes of $22''$ and $11''$ and an rms of $\sim 400\,\mathrm{\mu Jy\,beam^{-1}}$.  In Section~\ref{sec:spec_index} we combine these VLA observations at 344 MHz with the LOFAR observations at 144 MHz to measure spectral indices for the jellyfish galaxies identified in this work.}

\subsection{Perseus Cluster Galaxy Sample} \label{sec:gal_sample}

\begin{figure}
    \centering
    \includegraphics[width=\columnwidth]{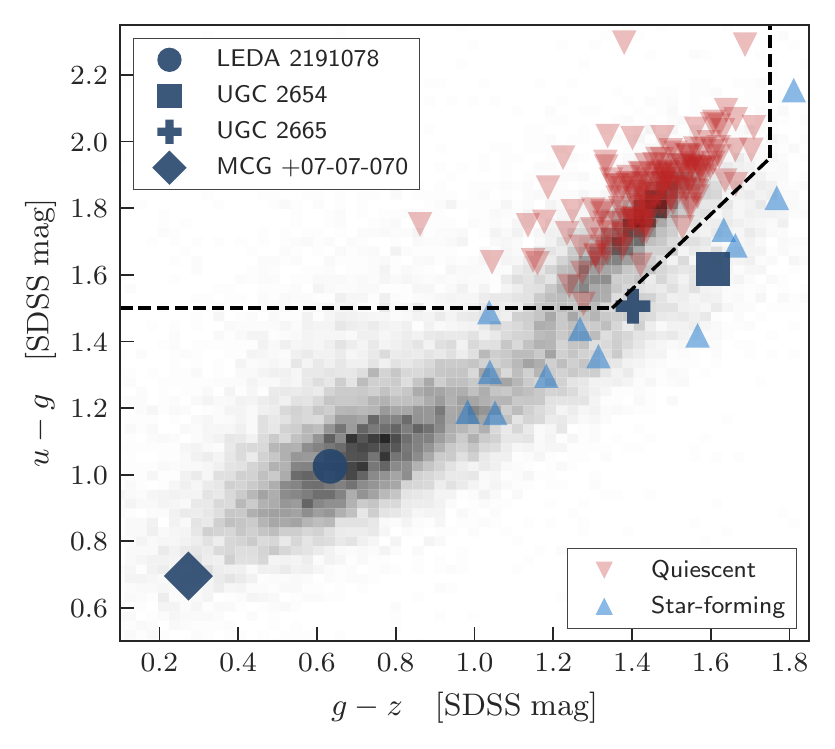}
    \caption{$ugz$ color-color diagram. Background grayscale shows distribution for all SDSS galaxies with redshifts between 0.008 and 0.028 and the dashed line shows the dividing line between star-forming and quiescent galaxies. Perseus jellyfish galaxies are shown with the large dark blue, distinct markers. Star-forming and quiescent Perseus cluster galaxies are shown with the blue and red triangle markers respectively.}
    \label{fig:ugz_diagram}
\end{figure}

We took advantage of previous spectroscopic observations in the vicinity of the Perseus cluster in order to build a sample of spectroscopic Perseus member galaxies. We queried the NASA IPAC Extragalactic Database (NED) for all galaxies that have a projected offset from NGC 1275 of $<\!1^\circ$, have a measured spectroscopic redshift, and are offset from the redshift of the Perseus cluster (which we take to be $z_\mathrm{Perseus} = 0.0179$) by $<\!3000\,\mathrm{km\,s^{-1}}$. The search radius of $1^\circ$ was chosen to ensure that all selected galaxies fall within the $\sim\!2^\circ \times 2^\circ$ field of the processed LOFAR image (which is centered on NGC 1275). The majority of these redshifts are from the Sloan Digital Sky Survey (SDSS), as the Perseus cluster was observed by the SDSS as a supplemental field \citep{adelman2007}, but we also included any other literature redshifts that are available from NED and satisfy our selection criteria. We then matched the selected galaxies with the SDSS DR16 \citep{ahumada2020} photometric catalog to obtain $ugriz$ magnitudes for each galaxy, keeping only galaxies with $m_r < 18$. For galaxy magnitudes we use extinction corrected SDSS model magnitudes. It is important to use extinction corrected magnitudes as the Perseus cluster is at small Galactic latitude and the field suffers from substantial reddening due to the Galactic foreground.  In total this procedure selected 166 spectroscopic members within $1^\circ$ of the Perseus cluster center.
\par
We separated quiescent and star-forming galaxies in this sample using an optical $u-g$ versus $g-z$ color-color diagram. This is a rough SDSS analogue to the commonly used $UVJ$ diagram \citep[e.g.,][]{wuyts2007}. In Fig.~\ref{fig:ugz_diagram} we plot $u-g$ color versus $g-z$ color. The background grayscale shows the distribution for all SDSS galaxies with redshifts between 0.008 and 0.028 (the same range as our Perseus member selection) and $m_r<18$. This full SDSS distribution is strongly bimodal and is used to define the boundary in the $ugz$ diagram separating quiescent and star-forming galaxies (dashed line in Fig.~\ref{fig:ugz_diagram}), which is determined by-eye. This color-color boundary is then used to classify Perseus member galaxies as star-forming or quiescent. The star-forming and quiescent Perseus members are shown in Fig.~\ref{fig:ugz_diagram} with the blue and red markers respectively.  Of the 166 Perseus cluster galaxies in our sample, 16 are identified as star-forming and 150 as quiescent.

\subsubsection{Identifying LOFAR jellyfish galaxies} \label{sec:id_jellyfish}

\begin{figure*}
    \centering
    \includegraphics[width=0.93\textwidth]{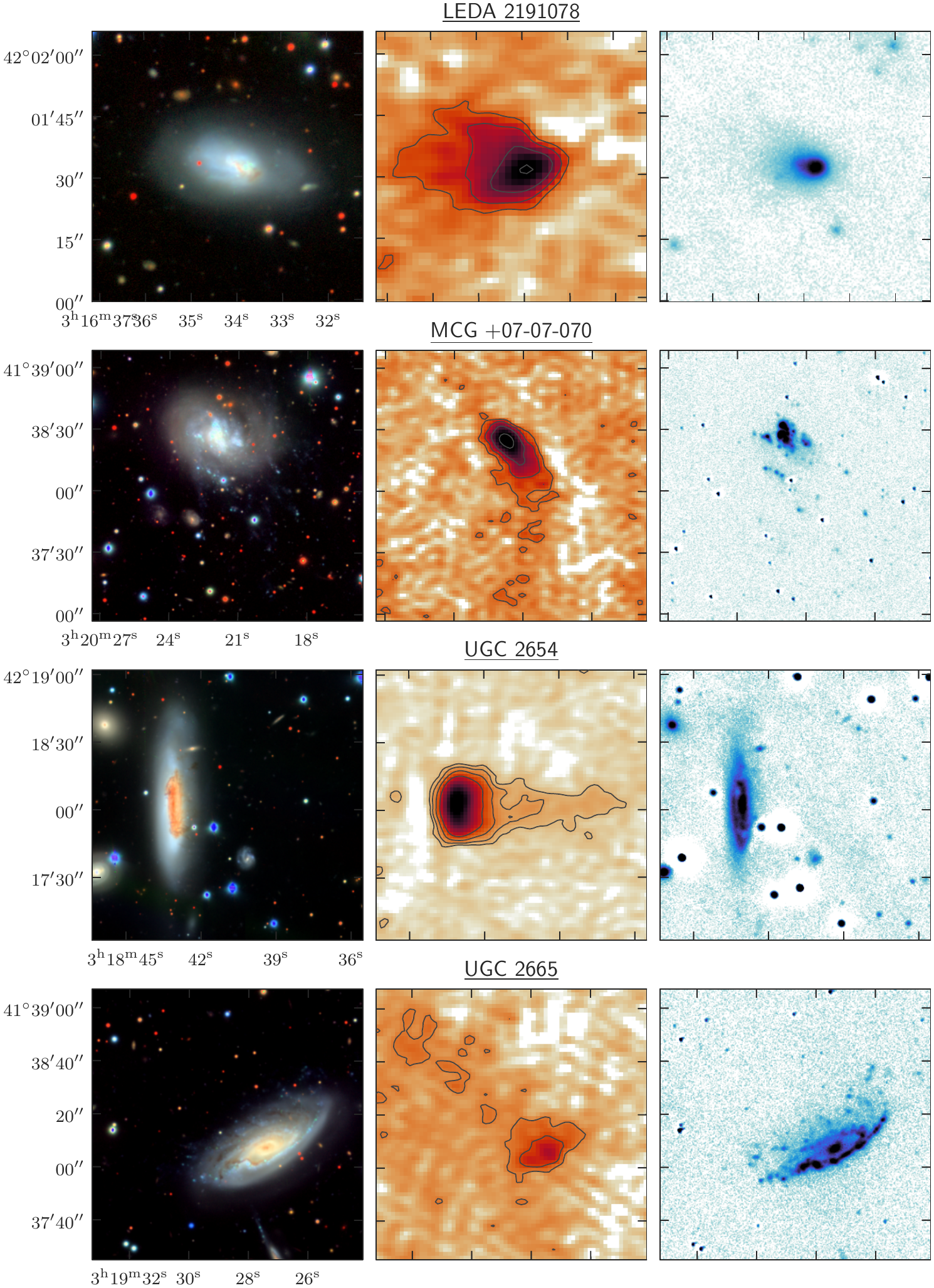}
    \caption{Panel images, highlighting the identified jellyfish galaxies at multiple wavelengths. Left-hand column shows $grz$ optical images from Subaru HSC. Middle column shows LOFAR $144\,\mathrm{MHz}$ images, with contours shown at the $\{3\sigma,\,6\sigma,\,12\sigma,\,\ldots\}$ level ($\sigma=100\,\mathrm{\mu Jy/beam}$ for the LOFAR image) and the $6''$ LOFAR beam shown in the lower left. Right-hand column shows the INT narrowband $\mathrm{H\alpha + [N\textsc{ii}]}$ image if available, and otherwise the SDSS $u$-band image.}
    \label{fig:panel_imgs}
\end{figure*}

To identify LOFAR jellyfish galaxies in the Perseus cluster we followed the selection criteria used by \citet{roberts2021_LOFARclust} and \citet{roberts2021_LOFARgrp} to identify jellyfish galaxies in low-redshift SDSS groups and clusters. This amounts to visually inspecting rest-frame optical and LOFAR $144\,\mathrm{MHz}$ images for all star-forming galaxies in Perseus. For optical imaging we use archival $g$-, $r$-, and $z$-band Subaru imaging of the Perseus cluster obtained from the Hyper Suprime-Cam (HSC) Legacy Archive \citep{tanaka2021}.  Jellyfish galaxies are then identified as those star-forming galaxies with $144\,\mathrm{MHz}$ emission that has an extended, one-sided asymmetry (i.e.,\ a tail) with respect to the galaxy stellar distribution (as traced by the rest-frame optical image). In our galaxy sample we have 16 star-forming galaxies, and through these visual inspection we identified four jellyfish galaxies with extended LOFAR radio continuum tails -- LEDA 2191078, MCG +07-07-070, UGC 2654, and UGC 2665. The position of these galaxies in Fig.~\ref{fig:ugz_diagram} are highlighted with the large, dark blue markers and in Table~\ref{tab:galaxy_samples} we summarize the general properties of these galaxies. In Fig.~\ref{fig:panel_imgs} we show panel images for each jellyfish galaxy with the $grz$ HSC image, the LOFAR $144\,\mathrm{MHz}$ image, and the INT $\mathrm{H\alpha+[N\textsc{ii}]}$ image (see Section~\ref{sec:INT_imaging}).

\subsection{Isaac Newton Telescope $\mathrm{H\alpha + [N\textsc{ii}]}$ imaging}
\label{sec:INT_imaging}

We also made use of public archival and newly obtained $\mathrm{H\alpha + [N\textsc{ii}]}$ narrowband and $r$ broad-band imaging of the Perseus cluster from the wide-field camera (WFC) on the Isaac Newton Telescope (INT)\footnote{http://casu.ast.cam.ac.uk/casuadc/ingarch/query}.  Archival imaging was available for MCG +07-07-070 \& UGC 2665. We have also obtained $\mathrm{H\alpha + [N\textsc{ii}]}$ and $r$-band imaging of UGC 2654 and LEDA 2191078 as part of the INT observing program ING.NL.21B.001 (PI Roberts). Narrowband observations were made with the \emph{WFCED337} filter at $665.7\,\mathrm{nm}$ and the \emph{WFCS6725} filter at $672.5\,\mathrm{nm}$ and broad-band observations were made with the \emph{WFCSloanR} filter. For the archival data, the narrowband integration time per target was $3600\,\mathrm{s}$ and the broad-band intergration time was $900\,\mathrm{s}$.  For the newly obtained imaging for UGC 2654 and LEDA 2191078, the narrowband integration time per target was $10800\,\mathrm{s}$ and the broad-band intergration time was $600\,\mathrm{s}$.
\par
Raw images were reduced in a standard fashion with the software package \textsc{theli} \citep{schirmer2013}, including bias correction and flat-fielding. GAIA DR2 \citep{gaia2016,gaia_dr2018} was used as the astrometric reference, PAN-STARRS photometry of unsaturated stars in the field was used for photometric calibration, and all individual images were background subtracted and then co-added into a single mosaic. Finally, the $r$-band images were used to subtract the continuum from the narrowband mosaics, creating continuum-subtracted $\mathrm{H\alpha + [N\textsc{ii}]}$ images.  The resulting image quality (IQ) in the $r$-band (narrowband) is $1.5''$ ($1.3''$) for MCG +07-07-070 and UGC 2665, $2.1''$ ($2.1''$) for LEDA 2191078, and $3.1''$ ($2.0''$) for UGC 2654.  In the continuum subtracted $\mathrm{H\alpha + [N\textsc{ii}]}$ images for MCG +07-07-070, UGC 2665, LEDA 2191078, and UGC 2654 we reach an rms level of 2.2, 2.5, 1.6, and 2.0 $\times 10^{-17}\,\mathrm{erg\,s^{-1}\,cm^{-2}\,arcsec^{-2}}$ respectively.
\par
\textcolor{black}{In Table~\ref{tab:galaxy_samples} we list the $\mathrm{H\alpha + [N\textsc{ii}]}$ fluxes for these four galaxies.  Fluxes are measured within $r_{25}$ (the 25th magnitude isophotal radius, obtained from the \textsc{hyperleda}\footnote{http://leda.univ-lyon1.fr/} database, \citealt{makarov2014}).  The flux uncertainties quoted in Table~\ref{tab:galaxy_samples} are a combination of three different sources, the statistical rms on individual pixels, the variation of the background level across the galaxy area, and the uncertainty on the photometric calibration.  For the pixel-by-pixel random error ($\sigma_\mathrm{px}$) we take a Monte Carlo approach and randomly shift the value of each pixel according to a Gaussian distribution with $\mu=0$ and $\sigma=rms$ (where $rms$ is the background rms measured over the image).  We then re-compute the galaxy flux from the randomly shifted pixel values.  This process is repeated 1000 times and we take $\sigma_\mathrm{px}$ to be equal to the standard deviation of the 1000 randomly shifted fluxes.  To gauge the potential variation of the background level over the area of the galaxy, $\sigma_\mathrm{bkg}$, we measure the background level over a number of regions around the galaxy that cover a similar area on the sky as the galaxy itself (this is following \citealt{boselli2015}).  We find that within these background regions there are fluctuations at the level of $\sim\!15$\% of the image rms.  Therefore we take $\sigma_\mathrm{bkg} = 0.15 \times rms \times A_\mathrm{px}$, where $A_\mathrm{px}$ is the number of pixels within $r_{25}$.  Finally, we assume a photometric calibration uncertainty, $\sigma_\mathrm{cal}$, of 10\% which is consistent with the scatter that we find in our zero-point estimation.  The uncertainty on the $\mathrm{H\alpha + [N\textsc{ii}]}$ flux, $\sigma_\mathrm{H\alpha + [N\textsc{ii}]}$, is then given as}
\begin{equation}
    \textcolor{black}{\sigma_\mathrm{H\alpha + [N\textsc{ii}]} = \sqrt{\sigma_\mathrm{px}^2 + \sigma_\mathrm{bkg}^2 + \sigma_\mathrm{cal}^2}}.
\end{equation}
\noindent
\textcolor{black}{For these data, $\sigma_\mathrm{bkg}$ and $\sigma_\mathrm{cal}$ are comparable and $\sigma_\mathrm{px}$ is a negligible contribution to the error on the total flux.}

\begin{table*}
    \centering
    \caption{Perseus cluster jellyfish galaxies.}
    \resizebox{\textwidth}{!}{
    \begin{threeparttable}
    \begin{tabular}{l c c c c c c c c c}
        \toprule
        Name & RA\tnote{a} & Dec\tnote{a} & Redshift\tnote{a} & \textcolor{black}{$M_\mathrm{star}$\tnote{b}} & \textcolor{black}{$\mathrm{SFR}$\tnote{b}} & $R / R_{500}$\tnote{c} & $\Delta v$\tnote{d} & $f_\mathrm{144\,MHz}$\tnote{e} & \textcolor{black}{$f_\mathrm{H\alpha+[NII]}$}\\
        & [deg] & [deg] & & \textcolor{black}{[$\mathrm{10^9\,M_\odot}$]} & \textcolor{black}{[$\mathrm{M_\odot\,yr^{-1}}$]} & & [$\mathrm{km\,s^{-1}}$] & [mJy] & \textcolor{black}{[$10^{-13}\,\mathrm{erg\,s^{-1}\,cm^2}$]} \\
        \midrule
        LEDA 2191078 & 49.14133 & 42.02572 & 0.0136 & \textcolor{black}{$3.7_{-0.8}^{+1.0}$} & \textcolor{black}{$0.4_{-0.2}^{+0.3}$} & 0.80 & 1289 & $19.8 \pm 3.6$ & $0.82 \pm 0.08$ \\[0.5em]
        MCG +07-07-070 & 50.09175 & 41.64072 & 0.0125 & \textcolor{black}{$3.5_{-0.7}^{+0.9}$} & \textcolor{black}{$0.7_{-0.3}^{+0.4}$} & 0.17 & 1618 & $40.8 \pm 5.7$ & \textcolor{black}{$1.92 \pm 0.19$} \\[0.5em]
        UGC 2654 & 49.67958 & 42.30058 & 0.0193 & \textcolor{black}{$93.7_{-19.3}^{+24.3}$} & \textcolor{black}{$2.8_{-1.0}^{+1.6}$} & 0.82 & 427 & $109.4 \pm 12.7$ & $1.48 \pm 0.16$ \\[0.5em]
        UGC 2665 & 49.86404 & 41.63531 & 0.0260 & \textcolor{black}{$15.5_{-3.2}^{+4.0}$} & \textcolor{black}{$0.8_{-0.3}^{+0.5}$} & 0.14 & 2441 & $21.5 \pm 5.6$ & \textcolor{black}{$2.09 \pm 0.21$} \\[0.5em]
        \bottomrule
    \end{tabular}
    \begin{tablenotes}
    \item[a] NASA IPAC Extragalactic Database\\
    \item[b] Computed using calibrations from \citet{leroy2019}. Stellar masses are determined from a combination of \textit{WISE1} luminosity with a mass-to-light ratio set according to \textit{WISE1}-\textit{WISE4} color.  SFRs are determined from \textit{WISE4} luminosity. \\
    \item[c] Projected offset from NGC 1275, we take $R_{500}=1.3\,\mathrm{Mpc}$ \citep{urban2014}. \\
    \item[d] Velocity offset from the Perseus cluster redshift, assuming $z_\mathrm{Perseus} = 0.0179$. \\
    \item[e] Includes all emission within the $3\sigma$ contour. Uncertainties include a noise contribution and a 10\% LoTSS DR2 calibration uncertainty (Shimwell et al., submitted) added in quadrature.
    \end{tablenotes}
    \end{threeparttable}}
    \label{tab:galaxy_samples}
\end{table*}

\section{Comments on individual galaxies} \label{sec:individual_gal}

Here we provide a brief discussion of each of the identified jellyfish galaxies.  We focus on the morphologies of these galaxies at multiple wavelengths, as illustrated by the panel images in Fig.~\ref{fig:panel_imgs}, as well as any literature references to these galaxies.

\subsection{LEDA 2191078} \label{sec:LEDA2191078}

LEDA 2191078 is a low-mass galaxy ($M_\mathrm{star} \sim 10^9\,\mathrm{\odot}$) with optical colors typical of a normal star-forming galaxy (Fig.~\ref{fig:ugz_diagram}).  Within Perseus, LEDA 2191078 is located at an intermediate cluster-centric radius ($R_\mathrm{proj} = 1038\,\mathrm{kpc} = 0.58\,R_{200}$) and a relatively large velocity offset from the Perseus cluster redshift ($\Delta v = 1289\,\mathrm{km\,s^{-1}}$).
\par
In the radio continuum, LOFAR detects LEDA 2191078 with a peak flux of $3.1\,\mathrm{mJy\,beam^{-1}}$, an integrated flux of $19.8 \pm 3.6\,\mathrm{mJy}$, and reveals extended and asymmetric emission. On the western side the $144\,\mathrm{MHz}$ emission is truncated well within the optical disk of the galaxy, but the eastern side shows an extended radio continuum tail extending beyond the optical extent of the galaxy.
\par
The $\mathrm{H\alpha + [N\textsc{ii}]}$ image of LEDA 2191078 shows a qualitatively similar morphology to that of the $144\,\mathrm{MHz}$ radio continuum, namely truncated on the western side and asymmetrically extended to the east, however over a much smaller extent. A prominent $\mathrm{H\alpha + [N\textsc{ii}]}$ peak is apparent west of the galaxy center which is co-spatial with a bright blue peak in the HSC image. LEDA 2191078 is briefly discussed in the Appendix B of \citet{meusinger2020} where it is proposed that the asymmetric optical morphology and dual brightness peaks in the optical are a result of a late-stage major merger. Here we suggest that these features may instead be driven by ram pressure between LEDA 2191078 and the Perseus cluster ICM. In particular, the $\mathrm{H\alpha + [N\textsc{ii}]}$ brightness peak offset to the west of the galaxy center is consistent with strong star formation being induced by ram pressure along the leading edge (i.e.,\ opposite to the tail direction) of cluster galaxies.  This would be consistent with previous studies, both observations \citep[e.g.,][]{gavazzi2001,boselli2021} and simulations \citep[e.g.,][]{bekki2014,troncoso-iribarren2020}, that have reported enhanced star formation on the leading side of galaxies undergoing RPS.

\subsection{MCG +07-07-070} \label{sec:MCG+07-07-070}

MCG +07-07-070 is the bluest galaxy in our sample and is located very close to the center of the Perseus cluster, offset by just $220\,\mathrm{kpc} = 0.12\,R_{200}$, with a relatively large velocity offset of $\Delta v = 1618\,\mathrm{km\,s^{-1}}$. Given its small cluster-centric radius (i.e.,\ high local ICM density; e.g.,\ \citealt{churazov2003}) and moderate-to-large velocity offset, MCG +07-07-070 is potentially experiencing very strong ram pressure as it passes through the cluster core \textcolor{black}{-- though this cannot be known with full confidence given that the cluster centric radius is observed in projection.}
\par
MCG +07-07-070 is detected at $144\,\mathrm{MHz}$ with a peak flux of $3.4\,\mathrm{mJy\,beam^{-1}}$ and an integrated flux of $40.8 \pm 5.7\,\mathrm{mJy}$. The radio morphology for MCG +07-07-070 is similar to that of LEDA 2191078, with an edge to the northeast that is truncated well within the optical disk and an extended, asymmetric tail to the south that reaches beyond the optical galaxy extent.
\par
In the optical, MCG +07-07-070 shows a highly disturbed morphology. Like LEDA 2191078, multiple brightness peaks in the optical are present near the galaxy nucleus with the brightest peak being found offset from the galaxy center to the north (roughly opposite to the direction of the radio tail). Furthermore, there are clear peaks of blue emission visible in the three-color Subaru image to the south of MCG +07-07-070, co-spatial with the $144\,\mathrm{MHz}$ tail. These sources resemble the ``fireballs'' of stellar emission that have been previously identified in the tails of other jellyfish galaxies \citep{yoshida2008,smith2010,yoshida2012,kenney2014,jachym2019}, consistent with recent star formation in the tail of MCG +07-07-070. The narrowband $\mathrm{H\alpha + [N\textsc{ii}]}$ image also shows highly disturbed emission within the galaxy. In particular, there are four bright regions of $\mathrm{H\alpha + [N\textsc{ii}]}$ emission near the galaxy nucleus; one at the center of the galaxy with the other three located to the east, west, and north of the galaxy center. $\mathrm{H\alpha + [N\textsc{ii}]}$ emission is also present to the south, beyond the optical extent of the galaxy. This emission is compact in nature and mostly coincident with the fireballs seen in the optical continuum.

\subsection{UGC 2654} \label{sec:UGC2654}
 UGC 2654 is an edge-on disk galaxy that is located outside of the cluster core ($R_\mathrm{proj} = 1066\,\mathrm{kpc} = 0.59\,R_{200}$) but quite close to the Perseus cluster redshift ($\Delta v = 427\,\mathrm{km\,s^{-1}}$). Fig~\ref{fig:ugz_diagram} shows that UGC 2654 has optical colors that are consistent with a dusty star-forming galaxy and prominent dust lanes can be seen near the center of the galaxy in the three-color Subaru image (Fig.~\ref{fig:panel_imgs}, left-hand column).
 \par
UGC 2654 is the brightest of the four jellyfish galaxies at $144\,\mathrm{MHz}$, with a peak flux of $15.8\,\mathrm{mJy\,beam^{-1}}$ and an integrated flux of $109.4 \pm 12.7\,\mathrm{mJy}$. The radio morphology consists of a bright core that covers the inner-most $\sim\!5\,\mathrm{kpc}$ of UGC 2654 and a stripped tail with a projected length of $\sim\!20\,\mathrm{kpc}$ extending beyond the optical disk to the west.  We do not find any evidence for an $\mathrm{H\alpha + [N\textsc{ii}]}$ tail alongside the clear tail in the radio continuum.  This may indicate that there is no stripped $\mathrm{H\alpha + [N\textsc{ii}]}$ emission associated with UGC 2654, though we also note that there are multiple saturated stars directly west of UGC 2654 that leave residuals after the continuum subtraction making it difficult to strongly constrain any potential $\mathrm{H\alpha + [N\textsc{ii}]}$ emission directly west of the galaxy disk.
\par
The broadband optical imaging of UGC 2654 shows a smooth disk morphology with a prominent ring of dust near the center of the galaxy visible through extinction. Close inspection of the HSC image shows small streams and clumps of dust that extend asymmetrically toward the western edge of UGC 2654 toward the direction of the radio continuum tail. This is is evidence for the stripping of dust in UGC 2654, consistent with previous works that have reported evidence for dust stripping in other nearby cluster galaxies \citep{kenney2015,cramer2021}. The Subaru $g$-band and INT $\mathrm{H\alpha + [N\textsc{ii}]}$ imaging for UGC 2654 is clearly asymmetric about the north-south axis. This is most easily seen through the bright arc of $\mathrm{H\alpha + [N\textsc{ii}]}$ emission along the eastern side of the galaxy.  Again, this may be a result of ram pressure catalyzing strong star formation along the leading side of UGC 2654; however, given that we find evidence for dust stripping toward the west, this asymmetric $\mathrm{H\alpha + [N\textsc{ii}]}$ and $g$-band emission could also be related to the asymmetric dust distribution and therefore asymmetric dust extinction.

\subsection{UGC 2665} \label{UGC2665}
 UGC 2665 is a spiral galaxy located near the center of Perseus ($R_\mathrm{proj} = 183\,\mathrm{kpc} = 0.10\,R_{200}$) and at a very large velocity offset of $\Delta v = 2441\,\mathrm{km\,s^{-1}}$.  Similar to MCG +07-07-070, this combination of small cluster-centric radius and large velocity offset indicates that UGC 2665 is \textcolor{black}{potentially} experiencing intense ram pressure \textcolor{black}{(again, modulo projection uncertainties)}. The optical colors of UGC 2665 are consistent with dusty star formation, which is also consistent with the clear dust lanes visible in the three-color Subaru image.
\par
UGC 2665 is our least significant detection at $144\,\mathrm{MHz}$, however we still see evidence for continuum emission from the central region of the galaxy at the $6\sigma$ level and for emission from an extended tail to the northeast at the $3\sigma$ level. The peak flux for UGC 2665 at $144\,\mathrm{MHz}$ is $1.6\,\mathrm{mJy\,beam^{-1}}$ and the integrated flux is $21.5 \pm 5.6\,\mathrm{mJy}$. \textcolor{black}{We note that UGC 2665 is spatially coincident with the northern edge of the 144 MHz emission from the central mini-halo around NGC 1275.  This makes it difficult to unambiguously associate the 144 MHz emission shown in Fig.~\ref{fig:panel_imgs} with a stripped tail given that we are likely observing a superposition of stripped material from UGC 2665 and low-frequency diffuse emission from the cluster center.} Given that the morphological signatures of RPS seen for UGC 2665 in the optical are consistent with the tail direction that we see in the radio continuum (see below) we believe that this tail emission is real, but the tail for UGC 2665 should still be considered a tentative detection.
\par
In the Subaru optical imaging and the INT narrowband $\mathrm{H\alpha + [N\textsc{ii}]}$, we see two morphological signatures of RPS for UGC 2665 that are consistent with the observed direction of the radio tail.  First, there is a bright arc of blue emission along the southwest edge of UGC 2665 that is visible in the Subaru $g$-band image, but most striking in the $\mathrm{H\alpha + [N\textsc{ii}]}$ image. This emission is again consistent with strong star formation being induced along the leading edge of the galaxy, opposite to the tail direction. Second, there is blue, knotty, extra-planar emission extending off of the disk to the north and to the east. This is visible in the Subaru $g$-band and the narrowband $\mathrm{H\alpha + [N\textsc{ii}]}$. UGC 2665 is also discussed in the Appendix B of \citet{meusinger2020} where it is suggested that the disturbed optical morphology of UGC 2665 may be a result of an interaction with the small galaxy to the south with a bridge of blue emission spanning between the two galaxies in projection (see Subaru image in Fig.~\ref{fig:panel_imgs}). We cannot rule out this possibility, and of course the observed morphology of UGC 2665 may be a result of both hydrodynamic interactions through ram pressure and dynamical interactions with galaxy neighbors.  Further constraints on this will require knowing the redshift of this small galaxy to the south, in order to determine whether it is part of the Perseus cluster or in the background. A spectroscopic redshift does not appear to have been measured to date and the SDSS photometric redshift for this source is $0.088 \pm 0.030$. The maximum likelihood photometric redshift is consistent with a background source, but it is also not well constrained and the source is still consistent with the redshift of Perseus at $\sim\!2-3\sigma$. This small galaxy to the south is detected in the narrowband $\mathrm{H\alpha + [N\textsc{ii}]}$ image after continuum subtraction, which does suggest that it is near the Perseus redshift.

\section{\textcolor{black}{Spectral Index Measurements}} \label{sec:spec_index}

\begin{figure*}
    \centering
    \includegraphics[width=\textwidth]{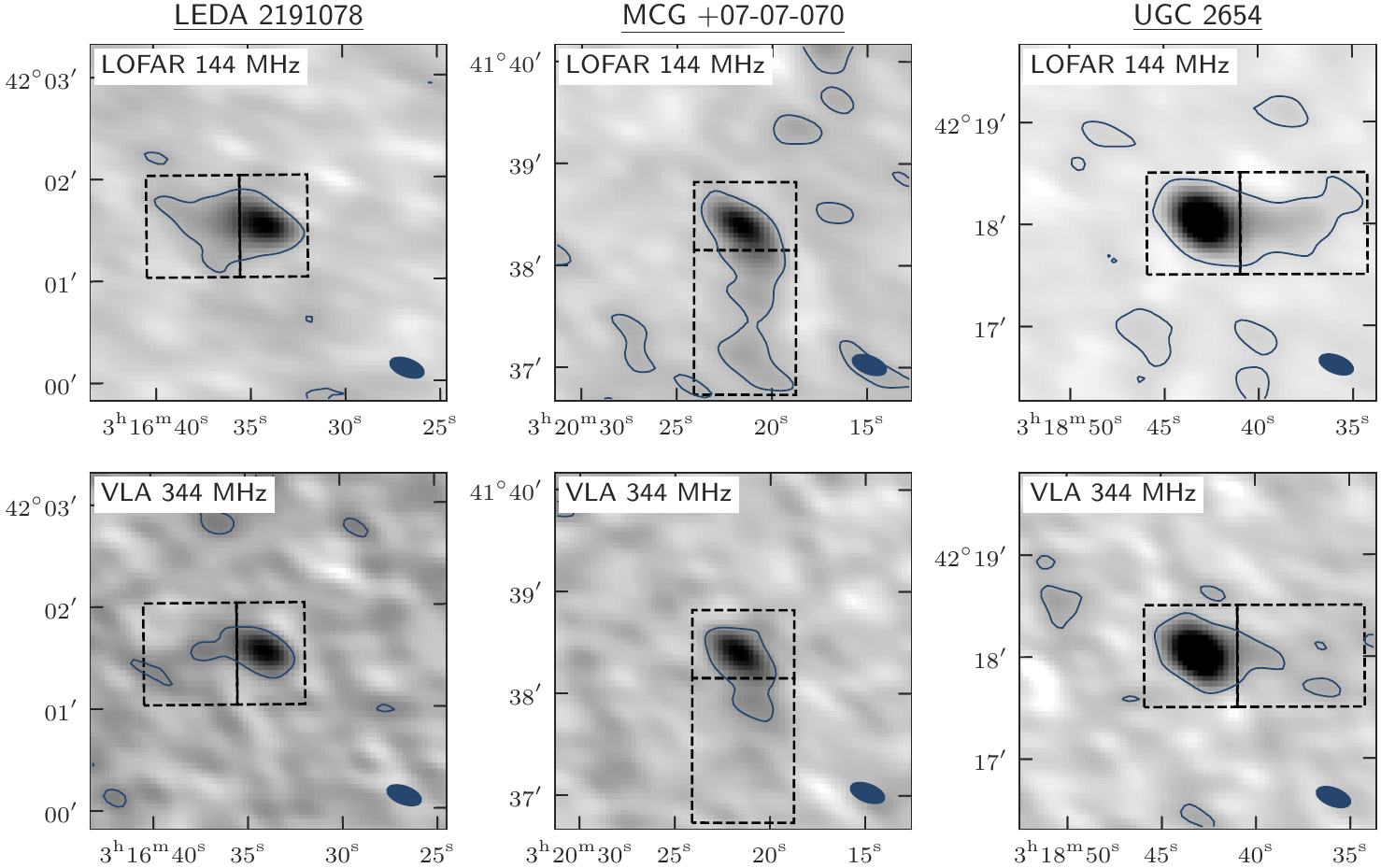}
    \caption{\textcolor{black}{Low resolution LOFAR 144 MHz cutouts (top row) and VLA 344 MHz cutouts (bottom row) of LEDA 2191078, MCG +07-07-070, and UGC 2654.  The dashed boxes show the regions used to measure the ``galaxy'' and ``tail'' spectral indices. The $22'' \times 11''$ ($7.5\,\mathrm{kpc} \times 3.7\,\mathrm{kpc}$) beam is shown in the lower right.}}
    \label{fig:lofar_vla_imgs}
\end{figure*}

\begin{figure}
    \centering
    \includegraphics[width=\columnwidth]{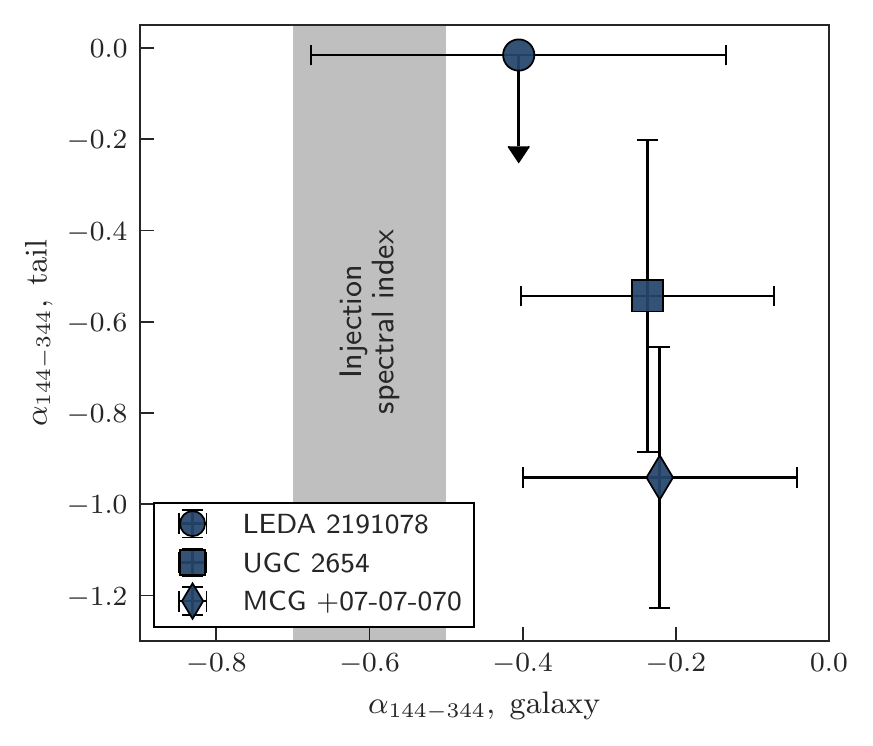}
    \caption{\textcolor{black}{Spectral indices measured between 144 MHz and 344 MHz for Perseus jellyfish galaxies.  The x-axis corresponds to a spectral index measured over the galaxy disk and the y-axis corresponds to a spectral index measured over the stripped tail. We also show the expected star-forming galaxy injection spectral index of $-0.7$ to $-0.5$ with the shaded region.  We do not measure a spectral index for UGC 2665 due to the blending of both the galaxy and tail emission with the central Perseus mini-halo in the smoothed 144 MHz image.}}
    \label{fig:spec_index}
\end{figure}

\textcolor{black}{Given the LOFAR imaging at 144 MHz and the VLA imaging at 344 MHz, we can explore the spectral indices of these jellyfish galaxies, both over the galaxy disk and over the stripped tail. To date there are relatively few spectral index measurements for RPS tails in jellyfish galaxies \citep[e.g.,][]{vollmer2004,vollmer2009,vollmer2013,chen2020,muller2021,ignesti2021}, especially extending to sub-GHz frequencies.  Constraining these spectral indices is critical for understanding the origins of the synchrotron emission in the stripped tails.  For example, if we are observing CREs produced in the galaxy disk and then subsequently removed from the galaxy via ram pressure, then a steepening of the spectral index along the tail should be observed due to synchrotron aging.  If substantial star formation is occurring within the stripped tail, this may also manifest itself via flattening in the measured spectral index.}
\par
\textcolor{black}{Before measuring spectral indices, we first match the resolution and pixel scale of the LOFAR and VLA imaging.  The LOFAR imaging has the higher resolution ($\sim 7'' \times 5''$ beam versus $\sim 22'' \times 11''$ beam), thus we smooth the LOFAR image to match the VLA beam shape and position angle. We also increase the pixel size of the LOFAR image by a factor of two in order to match the $3''$ pixels of the VLA image.  In Fig.~\ref{fig:lofar_vla_imgs} we show cutouts around LEDA 2191078, MCG +07-07-070, and UGC 2654 from the matched LOFAR and VLA images.  We do not include UGC 2665 in the spectral index analysis due to the blending of 144 MHz emission from UGC 2665 with emission from the Perseus mini-halo.  This was briefly mentioned in Section~\ref{sec:individual_gal} and this blending becomes even more substantial after smoothing the LOFAR image to match the VLA resolution.  As a result of this we cannot accurately measure a spectral index for UGC 2665 or its stripped tail since at this lower resolution we cannot disentangle the emission intrinsic to UGC 2665 and the emission intrinsic to the Perseus mini-halo.  In the VLA image, 344 MHz emission is detected over the galaxy disk of UGC 2665 but we do not detect emission at 344 MHz over the region of the inferred stripped tail.}
\par
\textcolor{black}{For each of the three jellyfish galaxies considered in this section, we measure two spectral index values between 144 MHz and 344 MHz; one over a region covering the main galaxy disk and one over a region covering the stripped tail.  We set these regions by eye using the low resolution LOFAR images (Fig.~\ref{fig:lofar_vla_imgs}, top row) and then measure flux densities from the LOFAR and VLA images within the same regions.  The ``galaxy'' and ``tail'' regions for each galaxy are shown in Fig.~\ref{fig:lofar_vla_imgs} with the dashed boxes. For consistency, we measure fluxes at both 144 MHz and 344 MHz over the entire boxes shown in Fig.~\ref{fig:lofar_vla_imgs}, even though the 344 MHz emission for the tails doesn't span the full area. Spectral indices are then calculated as:}
\textcolor{black}{
\begin{equation}
    \alpha_{\nu_1-\nu_2} = \frac{\log (S_1/S_2)}{\log (\nu_1/\nu_2)} \pm \frac{1}{\log (\nu_1/\nu_2)} \sqrt{\left(\frac{\sigma_1}{S_1}\right)^2 + \left(\frac{\sigma_2}{S_2}\right)^2},
\end{equation}
}
\noindent
\textcolor{black}{where $\nu_1$ and $\nu_2$ are the two frequencies that the spectral index is measured between, $S_1$ and $S_2$ are the flux densities at these frequencies, and $\sigma_1$ and $\sigma_2$ are the uncertainties on the flux density. To determine the uncertainties on the measured flux densities, $\sigma_1$ and $\sigma_2$, we add in quadrature a random component resulting from the noise in the images and a calibration uncertainty of 10\% for both the LOFAR and VLA fluxes.}
\par
\begin{table*}
    \centering
    {\color{black}
    \caption{Spectral index measurements}
    \begin{tabular}{l c c c c c c}
        \toprule
        Name & $S_\mathrm{144,\,galaxy}$ & $S_\mathrm{144,\,tail}$ & $S_\mathrm{344,\,galaxy}$ & $S_\mathrm{344,\,tail}$ & $\alpha_\mathrm{galaxy}$ & $\alpha_\mathrm{tail}$ \\
        & $[\mathrm{mJy}]$ & $[\mathrm{mJy}]$ & $[\mathrm{mJy}]$ & $[\mathrm{mJy}]$ && \\
        \midrule
        LEDA 2191078 & $13.9 \pm 1.4$ & $6.7 \pm 2.1$ & $9.8 \pm 2.1$ & $< 6.7$ & $-0.41 \pm 0.27$ & $< -0.02$ \\
        MCG +07-07-070 & $23.0 \pm 2.3$ & $23.5 \pm 3.3$ & $19.0 \pm 2.3$ & $10.3 \pm 2.1$ & $-0.22 \pm 0.18$ & $-0.94 \pm 0.29$ \\
        UGC 2654 & $86.5 \pm 8.6$ & $16.3 \pm 2.8$ & $70.3 \pm 7.3$ & $10.1 \pm 2.5$ & $-0.24 \pm 0.17$ & $-0.54 \pm 0.34$ \\
        \bottomrule
    \end{tabular}
    }
    \label{tab:spec_index}
\end{table*}

\textcolor{black}{In Table~\ref{tab:spec_index} we list the galaxy and tail flux densities at 144 MHz and 344 MHz as well as the resulting spectral indices.  For LEDA 2191078 we do not detect significant 344 MHz emission within the tail region from the VLA image, therefore in Table~\ref{tab:spec_index} we quote $3\sigma$ upper limits on the flux and spectral index. In Fig.~\ref{fig:spec_index} we plot the tail spectral index versus the galaxy spectral index for each of the three galaxies. It is clear from Fig.~\ref{fig:spec_index} that the spectral indices measured over the tail region are steeper than the spectral indices measured over the galaxy disk. The galaxy spectral indices are quite flat, ranging between -0.4 and -0.2. This is similar to the spectral index of -0.35 (between 144 MHz and 1.4 GHz) measured over the disk of the jellyfish galaxy JW100 \citep{ignesti2021}. These galaxy spectral indices are consistent with flattening due to ionization losses \citep{basu2015,chyzy2018} in a high density ISM. This flattening is expected at the low frequencies that we probe in this work and becomes increasingly important in the highest density regions of the ISM \citep{basu2015}, which is also consistent with enhanced ISM densities in jellyfish galaxies due to compression from ram pressure \citep[e.g.,][]{schulz2001,bekki2014,troncoso-iribarren2020,cramer2021}}.
\par
\textcolor{black}{In the stripped tail regions the spectral indices are not well constrained. For LEDA 2191078 we do not detect significant 344 MHz emission in the tail, and due to the relatively high noise level in the VLA image compared to the LOFAR image, the $3\sigma$ limit on the spectral index ($\alpha < -0.02$) that we derive does not meaningfully restrict the range of possible values.  For MCG +07-07-070 and UGC 2654 we do measure significant tail emission at both 144 MHz and 344 MHz and therefore are able to derive spectral indices. That said, this emission is still very faint (especially at 344 MHz) and therefore the error bars on the tail spectral indices are large.  We are not able to constrain whether there is steepening of the spectral index in the tail region due to synchrotron aging, as has been observed previously \citep{vollmer2004,chen2020,ignesti2021,muller2021}.  Within the uncertainties, the tail spectral indices for MCG +07-07-070 and UGC 2654 are consistent with both the expected injection spectral index for star-forming galaxies ($\alpha \sim -0.7$ to $-0.5$) and with moderate steepening in the tail region ($\alpha \sim -1$).}

\section{\textcolor{black}{Compact $\mathrm{H\alpha + [N\textsc{ii}]}$ Sources}}
\label{sec:Ha_srcs}

\begin{figure*}
    \centering
    \includegraphics[width=0.9\textwidth]{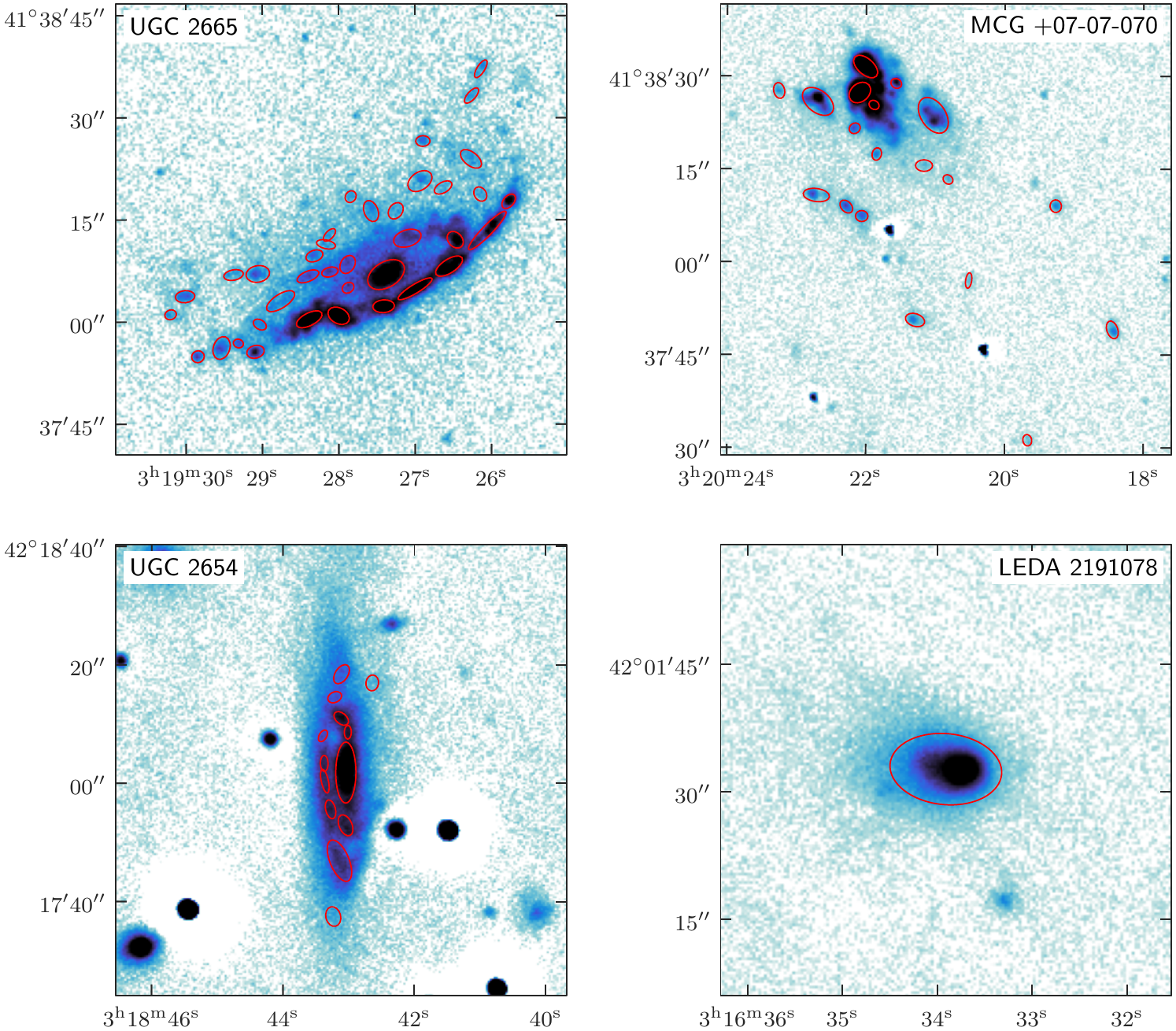}
    \caption{\textcolor{black}{Continuum subtracted $\mathrm{H\alpha + [N\textsc{ii}]}$ images of the four Perseus jellyfish galaxies.  In red we highlight the individual $\mathrm{H\alpha + [N\textsc{ii}]}$ sources that are identified through a dendrogram decomposition via the code \texttt{astrodendro}. The semi-major and semi-minor length for each red ellipse is given by $2\times \sigma_\mathrm{maj}$ and $2\times \sigma_\mathrm{min}$ from \texttt{astrodendro}.}}
    \label{fig:Ha_src_img}
\end{figure*}

\textcolor{black}{Given the relative proximity of the Perseus cluster, the typical seeing of the INT observations used in this work ($\sim\!2''$) traces scales on the order of hundreds of pc.  While this does not quite reach the typical size of \textsc{Hii} regions ($\lesssim \! 100\,\mathrm{pc}$, \citealt{kennicutt1979}), these subkiloparsec scales allow us to probe $\mathrm{H\alpha + [N\textsc{ii}]}$ sources near to the scale of individual star-forming regions with the INT imaging.  In this section we systematically identify such sources and quantify their properties, in particular with respect to the inferred direction of RPS.}

\subsection{\textcolor{black}{Dendrogram Source Identification}} \label{sec:dendro}

\textcolor{black}{We identified these bright $\mathrm{H\alpha + [N\textsc{ii}]}$ sources systematically with the \textsc{python} package \texttt{astrodendro}\footnote{http://www.dendrograms.org/} \citep{robitaille2019} that decomposes the $\mathrm{H\alpha + [N\textsc{ii}]}$ galaxy image into dendrograms. The resulting dendrogram is a tree representing hierarchical structure in the galaxy image. This consists of structures that are split into smaller substructures (``branches'') and structures with no substructures (``leaves''). We identified $\mathrm{H\alpha + [N\textsc{ii}]}$ sources in each galaxy simply by taking the leaves of the dendrogram for each galaxy. We set the minimum value considered in the dendrogram to be $3\times$ the background rms, ensuring that these sources are detected above the background. We set the minimum significance for each leaf to be $1\times \mathrm{rms}$, this means that neighboring leaves that differ by less than $1\times \mathrm{rms}$ are merged into a single structure.  Finally, we use the $\mathrm{H\alpha + [N\textsc{ii}]}$ image quality to set the minimum number of pixels required for an independent source.  Specifically, we set the minimum source area to the area of a circle with diameter equal to the FWHM seeing of the INT observations. With these parameters, we compute dendrograms for the $\mathrm{H\alpha + [N\textsc{ii}]}$ images of the four jellyfish galaxies and take $\mathrm{H\alpha + [N\textsc{ii}]}$ sources to be the leaves of each dendrogram. This process identifies 37 $\mathrm{H\alpha + [N\textsc{ii}]}$ sources for UGC 2665, 20 $\mathrm{H\alpha + [N\textsc{ii}]}$ sources for MCG +07-07-070, 13 $\mathrm{H\alpha + [N\textsc{ii}]}$ sources for UGC 2654, and 1 $\mathrm{H\alpha + [N\textsc{ii}]}$ source for LEDA 2191078.  We note that we exclude sources by hand identified by \texttt{astrodendro} that are associated with residuals from foreground stars after continuum subtraction. We highlight the identified sources in Fig.~\ref{fig:Ha_src_img}, the majority of which are found within the stellar disk of each galaxy (in projection) but we do identify a smaller number of $\mathrm{H\alpha + [N\textsc{ii}]}$ sources beyond the galaxy optical radius along the direction of the stripped tail; this is particularly true for MCG +07-07-070.}

\subsection{\textcolor{black}{$\mathrm{H\alpha + [N\textsc{ii}]}$ Sources Relative to Tail Direction}}
\label{sec:src_tail_direction}

\begin{figure}
    \centering
    \includegraphics[width=\columnwidth]{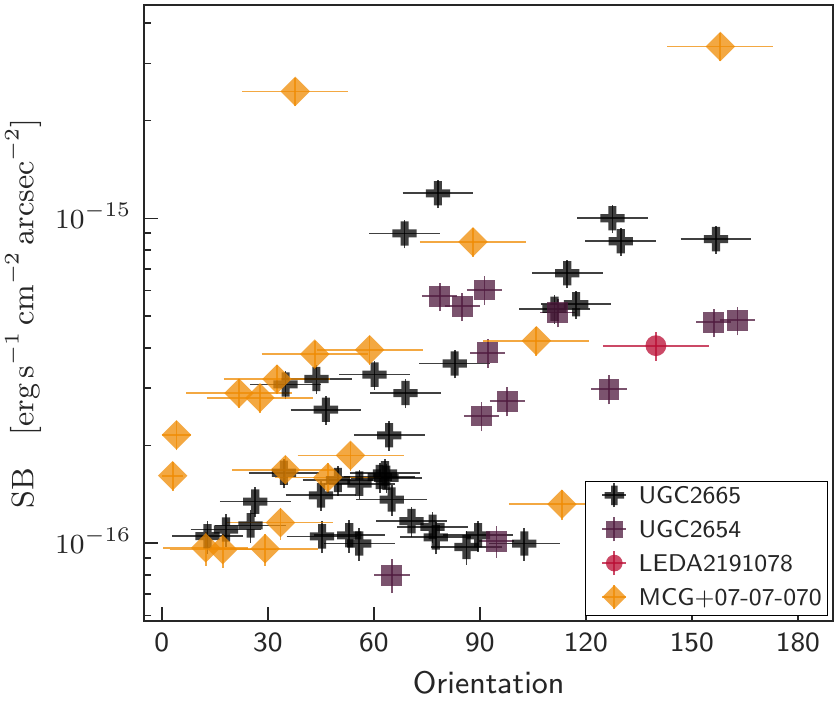}
    \caption{\textcolor{black}{Surface brightness of $\mathrm{H\alpha + [N\textsc{ii}]}$ sources versus orientation with respect to the radio continuum tail.  An angle of $0^\circ$ corresponds to a source along the direction of the stripped tail and an angle of $180^\circ$ corresponds to a source directly opposite to the direction of the stripped tail.  Error bars on the $\mathrm{H\alpha + [N\textsc{ii}]}$ surface brightness are calculated following the method outlined in Section~\ref{sec:INT_imaging} and error bars on the orientation angles are determined from visual determination of tail directions (see Section~\ref{sec:src_tail_direction} for details).}}
    \label{fig:Ha_src_ang}
\end{figure}

\textcolor{black}{In this section we explore the dependence of $\mathrm{H\alpha + [N\textsc{ii}]}$ source surface brightness with respect to the direction of the radio continuum tail for these Perseus jellyfish galaxies.  The direction of the stripped tail is estimated following \citet{roberts2021_LOFARclust,roberts2021_LOFARgrp}; namely, the tail direction is visually assigned a direction between $0^\circ$ and $360^\circ$ where $0^\circ = \mathrm{west}$ and $90^\circ = \mathrm{north}$.  Along with this ``best'' tail direction estimate, a range corresponding to the minimum and maximum ``plausible'' tail directions is also estimated which is used to quantify the uncertainty on these directions.  For MCG +07-07-070 a tail direction of $290 \pm 15$ degrees is estimated, $135 \pm 10$ degrees for UGC 2665, $0 \pm 5$ degrees for UGC 2654, and $165 \pm 15$ degrees for LEDA 2191078.}
\par
\textcolor{black}{In Fig.~\ref{fig:Ha_src_ang} we plot $\mathrm{H\alpha + [N\textsc{ii}]}$ source surface brightness versus the angular orientation of each source with respect to the direction of the radio continuum tail -- all measured relative to the galaxy center.  An orientation angle of $0^\circ$ corresponds to a source directly in the direction of the stripped tail and an orientation angle of $180^\circ$ corresponds to a source directly opposite to the direction of the stripped tail (i.e.,\ on the leading side). Any sources that overlap with the center of their host galaxy are excluded from Fig.~\ref{fig:Ha_src_ang}, as orientation angles are measured with respect to the galaxy center and therefore the orientation angle is not well defined in these cases.  We mark sources from each individual galaxy with a distinct color and marker shape (see Fig.~\ref{fig:Ha_src_ang} for details).  In Fig.~\ref{fig:Ha_src_ang} there is a clear positive correlation between $\mathrm{H\alpha + [N\textsc{ii}]}$ surface brightness and the orientation angle of sources with respect to the stripped tail.  This is confirmed quantitatively at high significance with Spearman's rank correlation test ($r_S=0.48,\;p_S \sim 10^{-5}$).  This suggests that $\mathrm{H\alpha + [N\textsc{ii}]}$ emission, and therefore star formation, is enhanced along the leading side of the jellyfish galaxies, consistent with gas compression induced by ram pressure along the galaxy-ICM interface.  As alluded to in Section~\ref{sec:individual_gal}, such an effect has been predicted by simulations of RPS \citep[e.g.,][]{bekki2014,troncoso-iribarren2020} and been reported observationally for a limited number of galaxies experiencing ram pressure \citep[e.g.,][]{gavazzi2001,boselli2021}.  Further work is required to test whether these leading-side star formation enhancements are commonplace for galaxies undergoing RPS, and the results reported here are a step in that direction.}
\par
\textcolor{black}{Finally, for MGC +07-07-070 we detect a number of $\mathrm{H\alpha + [N\textsc{ii}]}$ sources that are co-spatial with the stripped tail, beyond the optical extent of the galaxy.  The surface brightness of these sources is relatively low (they form the locus of points in Fig.~\ref{fig:Ha_src_ang} with low surface brightnesses and orientation angles $<\!60^\circ$), roughly an order of magnitude fainter than the brightest sources in MCG +07-07-070.  Assuming a standard prescription from \citet{kennicutt2012}, these tail sources would contribute a total $\mathrm{H\alpha}$ SFR of $\sim\!10^{-2}\,\mathrm{M_\odot\,yr^{-1}}$, ranging per source from $\sim\!5\times 10^{-4}\,\mathrm{M_\odot\,yr^{-1}}$ to $\sim\!4\times 10^{-3}\,\mathrm{M_\odot\,yr^{-1}}$.  We note that this is meant to be an order of magnitude estimate and does not include any correction for the [N\textsc{ii}] contribution to the narrowband flux, nor are they dust corrected.  Depending on the $\mathrm{H\alpha}$-to-[N\textsc{ii}] ratio for these sources the SFRs may be overestimated by a factor of two or so.}

\section{\textcolor{black}{Radio Luminosity versus SFR}}
\label{sec:radio_lum_sfr}

\begin{figure}
    \centering
    \includegraphics[width=\columnwidth]{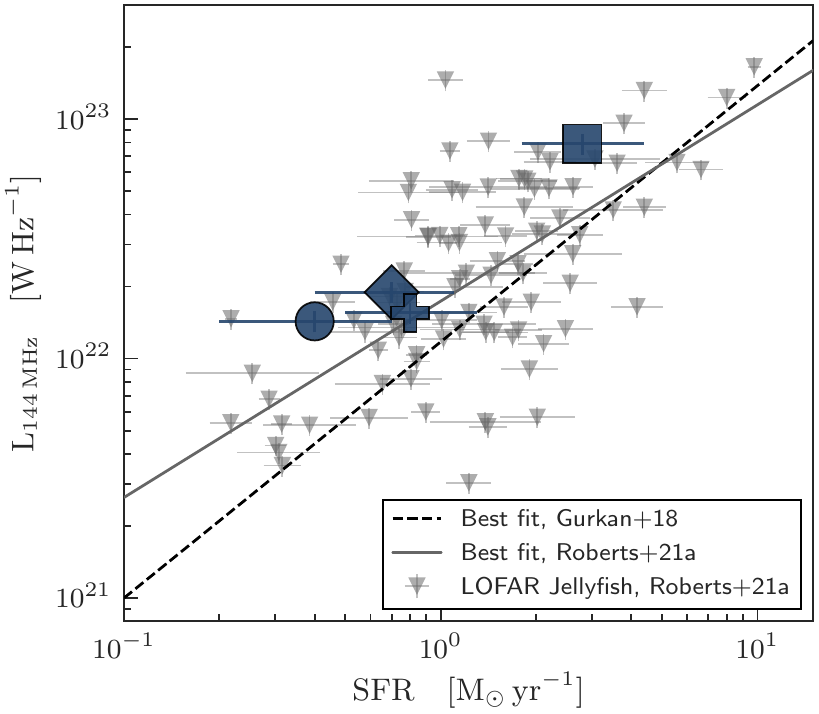}
    \caption{\textcolor{black}{Low frequency radio luminosity versus SFR relation.  LOFAR 144 MHz luminosity versus SFR for the Perseus jellyfish galaxies presented in this work (large blue symbols, same marker types as Fig.~\ref{fig:ugz_diagram}) and the larger low-$z$ LOFAR jellyfish galaxy sample from \citet{roberts2021_LOFARclust} (gray triangles).  We also show the best-fit powerlaw relationships for the jellyfish sample from \citet{roberts2021_LOFARclust} (solid gray) and the low-$z$ star-forming galaxy sample from \citet{gurkan2018} (dashed black).}}
    \label{fig:L144_sfr}
\end{figure}

\textcolor{black}{In \citet{roberts2021_LOFARclust} we reported a slight excess in 144 MHz luminosity for jellyfish galaxies relative to the 144 MHz luminosity - SFR relation for normal, low-$z$ star-forming galaxies.  In this section we test whether the jellyfish galaxies in the Perseus cluster show a similar systematic shift to large 144 MHz luminosities relative to their measured SFR.}

\subsection{\textcolor{black}{Determining SFRs}}

\textcolor{black}{We compute integrated SFRs for the four Perseus jellyfish galaxies following the procedure outlined in \citet{leroy2019}.  \citet{leroy2019} provide SFR calibrations based on UV (\textit{GALEX}) photometry, IR (\textit{WISE}) photometry, or a combination of the two.  All four of the jellyfish galaxies are covered by the \textit{WISE} All-Sky Survey, however only the central field of the Perseus cluster has been imaged by \textit{GALEX} and therefore two of the four jellyfish galaxies (LEDA 2191078 and UGC 2654) lack \textit{GALEX} imaging.  For this reason we compute SFRs using the \textit{WISE4}-only ($22\,\mathrm{\mu m}$) prescription from \citet{leroy2019}, so that the same calibration can be used for all four galaxies.}
\par
\textcolor{black}{\textit{WISE4} images are obtained for each galaxy from the unWISE database\footnote{http://unwise.me/} \citep{lang2016}, and all four jellyfish galaxies have strong detections ($\mathrm{S/N} \gtrsim 100$) at $22\,\mathrm{\mu m}$.  Following \citet{leroy2019}, the SFR is then given as}
\begin{equation}
    \textcolor{black}{\mathrm{SFR} = \nu L_{\nu,W4} \times 10^{-42.63}}
\end{equation}
\noindent
\textcolor{black}{where $\nu$ is the effective frequency of the \textit{WISE4} filter and $L_{\nu,W4}$ is the \textit{WISE4} specific luminosity in cgs units. To compute SFRs we use the total \textit{WISE4} luminosity within $2 \times r_{25}$, where we take $r_{25}$ for each galaxy from the \textsc{hyperleda} database \citep{makarov2014}.}

\subsection{\textcolor{black}{The $L_\mathrm{144\,MHz} - \mathrm{SFR}$ Relation}}

\textcolor{black}{In Fig.~\ref{fig:L144_sfr} we show the $L_\mathrm{144\,MHz} - \mathrm{SFR}$ relation.  The dashed line corresponds to the best-fit relation for normal, low-$z$ star-forming galaxies from \citet{gurkan2018}, the solid line shows the best-fit relation for the LOFAR jellyfish galaxies presented in \citet{roberts2021_LOFARclust}, the small markers correspond to the individual jellyfish galaxies from \citet{roberts2021_LOFARclust}, and the large blue markers correspond to the Perseus cluster jellyfish galaxies reported here. For all jellyfish galaxies, both from this work and from \citet{roberts2021_LOFARclust}, the 144 MHz luminosity includes continuum emission both from the galaxy disk and from the stripped tail.  The SFRs for the jellyfish galaxies in \citet{roberts2021_LOFARclust} are taken from the GSWLC-2\footnote{https://salims.pages.iu.edu/gswlc/} catalog \citep{salim2016,salim2018} and are derived from SED fitting using \textit{GALEX}, SDSS, and \textit{WISE} photometry, whereas the SFRs for the Perseus cluster jellyfish were calculated using a \textit{WISE4}-only prescription.  That said, the SFR prescriptions from \citet{leroy2019} are calibrated using SFRs from the GSWLC-2 catalog, therefore the SFRs in Fig.~\ref{fig:L144_sfr} should be generally comparable.  Perseus cluster galaxies are not included in the GSWLC-2 catalog (since Perseus is not part of the main SDSS survey), therefore a fully direct comparison is not possible.}
\par
\textcolor{black}{In Fig.~\ref{fig:L144_sfr} the four Perseus jellyfish galaxies are found clearly above the \citet{gurkan2018} relation, again consistent with jellyfish galaxies having excess emission at 144 MHz relative to normal star-forming galaxies. Relative to the \citet{gurkan2018} relation, the Perseus jellyfish have 144 MHz luminosities that are enhaced by a factor ranging from 1.7 to 3.2. We discuss potential explanations for this radio luminosity enhancement below in Sect.~\ref{sec:discussion_L144_sfr}}

\section{Discussion \& Conclusions} \label{sec:conclusion}

\textcolor{black}{In this paper we present the first examples of jellyfish galaxies in the Perseus cluster. These four galaxies are identified on the basis of extended, one-sided radio tails observed by LOFAR at $144\,\mathrm{MHz}$. While these jellyfish galaxies are identified from low-frequency radio observations, we show that their optical morphologies (both broad-band and narrowband $\mathrm{H\alpha + [N\textsc{ii}]}$) are consistent with clear perturbations resulting from ram pressure.  All four galaxies show some evidence of enhanced star formation along the leading side of the galaxy and have enhanced radio luminosities relative to the standard low-frequency $L_\mathrm{144\,MHz} - \mathrm{SFR}$ relation.  The radio spectral indices of these jellyfish galaxies are quite flat when measured over the galaxy disk and steeper over the stripped tails. Below we provide a brief discussion of this work and conclude with a summary of the primary results.}

\subsection{\textcolor{black}{Galaxy and Tail Spectral Indices}}\label{sec:discussion_spec_index}

\textcolor{black}{With 144 MHz and 344 MHz imaging we show that these Perseus jellyfish galaxies tend to have flat spectral indices over the galaxy disk region ($\alpha \sim -0.4$ to  $-0.2$) with steeper, though relatively unconstrained, spectral indices over the region of the stripped tail. A general trend of spectral steepening along RPS tails has been reported by previous studies (\citealt{vollmer2004,chen2020,ignesti2021,muller2021}; though we note that not all RPS tails have steep spectral indices, e.g., \citealt{vollmer2009}), which is consistent with synchrotron aging as CREs are removed from the galaxy. The uncertainties on the tail spectral indices in this work are too large to differentiate between the expected injection spectral index from star formation ($\alpha \sim -0.7$ to $-0.5$) and a spectral index indicative of synchrotron aging relative to this value ($\alpha \lesssim -1$).  The flat spectral indices within the disks of these galaxies are consistent with the disk spectral index ($\alpha = -0.35$) reported by \citet{ignesti2021} at low-frequencies for the jellyfish galaxy JW100.  Other studies have found steeper spectral indices over the disks of galaxies undergoing RPS, that are more in line with the $\alpha \sim -0.7$ to $-0.5$ typically found for star-forming galaxies \citep{vollmer2004,vollmer2013,muller2021}. An important difference is that this work and \citet{ignesti2021} probe down to low-frequencies (144 MHz at the low-frequency end), whereas the spectral indices measured in \citet{vollmer2004,vollmer2013} and \citet{muller2021} are measured at higher frequencies (from 1.4 GHz to 5 GHz and from 1.4 GHz to 2.7 GHz, respectively). This difference in spectral index at low- and high-frequencies may be pointing to physics in the ISM. For example, at high ISM densities ionization losses become increasingly important at low-frequencies and can lead to a flattening of the low-frequency spectral index \citep[e.g.,][]{basu2015,chyzy2018}.  High ISM densities are also expected in jellyfish galaxies due to compression induced by ram pressure \citep{schulz2001,bekki2014,troncoso-iribarren2020,cramer2021}. } 
\par
\textcolor{black}{The number of jellyfish galaxies with spectral index measurements at low-frequencies is currently too small to make any clear conclusions, but there are excellent prospects for increasing this sample size moving forward.  What will be particularly valuable is spectral index measurements across a range of frequencies that are resolved across the disks of jellyfish galaxies.  Since enhanced ISM densities are primarily expected on the leading side of jellyfish galaxies (i.e., opposite to the tail direction), this spectral index flattening due to ionization losses should also be strongest on the leading side. In Roberts et al. (in prep.), we will present low-frequency spectral index maps for the $\sim\!10$ Coma Cluster jellyfish galaxies with RPS tails detected at both 1.4 GHz in \citet{chen2020} and 144 MHz in \citet{roberts2021_LOFARclust}. This will allow us to explore the spatial variation of spectral index across the galaxy disks and stripped tails at $\sim 3\,\mathrm{kpc}$ resolution. Given its proximity, the Virgo Cluster will also be a key environment in order to probe very high physical resolutions.  High frequency radio continuum imaging is already available \citep[e.g.,][]{vollmer2004,vollmer2009,vollmer2013} and high quality LOFAR HBA imaging at 144 MHz will likely become available in the future.  Furthermore, given the VIVA \textsc{Hi} survey \citep{chung2009} and the VERTICO CO(2-1) survey \citep{brown2021}, a direct comparison between spectral index and local ISM density would be possible.}

\subsection{\textcolor{black}{An Offset from the $L_{144\,\mathrm{MHz}} - \mathrm{SFR}$ Relation for Jellyfish Galaxies}} \label{sec:discussion_L144_sfr}

\textcolor{black}{In \citet{roberts2021_LOFARclust} we report evidence for a small offset for a population of $\sim\!100$ jellyfish galaxies above the standard $L_{144\,\mathrm{MHz}} - \mathrm{SFR}$ relation. Similarly, in both the Coma and Virgo clusters, galaxies undergoing RPS are found systematically above the radio-infrared relation for normal star-forming galaxies \citep[here the radio luminosities are measured at 1.4 GHz,][]{murphy2009,vollmer2013,chen2020}. On an individual galaxy basis, a similar offset has been published by \citet{ignesti2021} for the jellyfish galaxy JW100 from the GASP survey \citep{poggianti2017}, and we also see such an offset for the four Perseus cluster jellyfish galaxies in this work.  Thus there is evidence that jellyfish galaxies populate a locus above the standard radio luminosity vs. SFR relation. The origin of this offset is still an open question, and in principle could be driven by sources of radio emission not associated with star formation, differences between the synchroton aging timescale and the timescale of SFR quenching, compression of the magnetic field, or some combination of the above.}
\par
\textcolor{black}{The SFRs in Fig.~\ref{fig:L144_sfr} are only measured over the main disk of the galaxy, whereas the 144 MHz luminosities include emission both from the galaxy disk and the stripped tail.  Therefore if there is substantial star formation occuring within the stripped tail, this is not accounted for in the estimated SFR, and thus any synchrotron emission associated with this extraplanar star formation would tend to offset galaxies above the $L_{144\,\mathrm{MHz}} - \mathrm{SFR}$ relation.  That said, the offsets observed in Fig.~\ref{fig:L144_sfr} for the Perseus jellyfish galaxies are roughly a factor of 2-3$\times$ above the \citet{gurkan2018} relation.  Given that the \citet{gurkan2018} relation is roughly linear between 144 MHz luminosity and SFR, it is very difficult to imagine enough star formation occuring in the tail to make up a SFR deficit of 2-3$\times$.  Indeed, for MCG +07-07-070 which does show evidence for ongoing star formation in the stripped tail, the $\mathrm{H\alpha + [N\textsc{ii}]}$ emission that is co-spatial with the stripped tail only makes up $\sim\!5\%$ of the total $\mathrm{H\alpha + [N\textsc{ii}]}$ flux.  Taking the $\mathrm{H\alpha + [N\textsc{ii}]}$ flux as a SFR tracer, this would not be nearly enough star formation in the tail to account for the offset seen in Fig.~\ref{fig:L144_sfr}.  It is also possible that emission from a central AGN in these galaxies could be contributing 144 MHz flux not associated with star formation. Three of the four jellyfish galaxies have measured SDSS spectra (all except UGC 2654).  According to the BPT criteria there is some evidence for AGN emission in UGC 2665, but no evidence for AGN emission in MCG +07-07-070 or LEDA 2191078.  There does not appear to be a publicly available rest-frame optical spectrum for UGC 2654 so we cannot comment on any evidence for AGN emission from optical line ratios in this case.}
\par
\textcolor{black}{An alternative explanation, as suggested by \citet{ignesti2021}, is that this offset is a product of jellyfish galaxies undergoing rapid star formation quenching.  RPS is associated with a rapid ($\lesssim\!1\,\mathrm{Gyr}$) shut-off of star formation \citep[e.g.,][]{quilis2000}, which can be due to both efficient stripping of star-forming gas and potentially rapid gas consumption due to temporarily enhanced SFRs.  In this sense the excess 144 MHz emission for jellyfish galaxies may originate from star formation which was higher in the past and is now being quenched.  At 144 MHz the synchrotron radiative timescale is $\sim\!200\,\mathrm{Myr}$ assuming a $1\,\mathrm{\mu G}$ magnetic field or $\sim\!25\,\mathrm{Myr}$ assuming a $10\,\mathrm{\mu G}$ magnetic field, which roughly corresponds to the timescale over which the observed radio emission is tracing star formation.  Therefore a reduction of star formation by a factor of 2-3$\times$ over the past 25-200 Myr would be consistent with the observed radio excess for Perseus jellyfish galaxies.  \citet{ignesti2021} perform a spectrophotometric star formation history (SFH) analysis of MUSE IFU data for the jellyfish galaxy JW100 and find evidence that the SFR in the disk of JW100 was indeed higher in the past at the level required to explain the observed radio excess.  Currently we lack the necessary data products to perform a resolved SFH analysis of the four jellyfish galaxies presented here, but future observations, for example with the forthcoming WEAVE-IFU instrument on the William Herschel Telescope, would permit such an analysis and enable the direct testing of this scenario.}
\par
\textcolor{black}{The enhanced radio emission may also be a product of perturbations to the galaxy magnetic field from ram pressure. For example, compression of the ISM and magnetic field due to ram pressure would lead to enhanced radio continuum emission.  This explanation has been invoked previously to explain excess radio continuum emission in star-forming cluster galaxies \citep[e.g.,][]{gavazzi1999,murphy2009,vollmer2013,chen2020}, and may be responsible for the observed radio excess for these Perseus jellyfish galaxies.  Resolved maps of SFR for these galaxies would help to further constrain this possibility by allowing us to spatially resolve the $L_\mathrm{144\,MHz} / \mathrm{SFR}$ ratio and test whether this ratio is globally or locally enhanced in these galaxies.  For example, compression of the magnetic field by ram pressure should be strongest on the leading edge, thus it is plausible that the radio excess is highest in this region of the galaxy.  That said, such compression of the ISM may also induce star formation, and we do show that all four of the Perseus jellyfish show evidence for enhanced star formation along the leading side of the galaxy.  Therefore it is also possible that on the leading side of these galaxies the radio luminosity and SFR are both enhanced in proportion, leaving the ratio relatively unchanged.  Lastly, as discussed above, compression of the ISM will also lead to higher ionization losses and in turn less radio continuum flux at low-frequencies.  Thus there may be competition between increased continuum emission due to a stronger magnetic field and decreased continuum emission due to these energy losses.  Predicting exactly what is expected observationally in this compression scenario is nontrivial. There are multiple relevant physical effects that could affect the galaxy SFR, magnetic field strength, and low-frequency radio continuum flux.}
\par
\textcolor{black}{The physical origin of this enhanced radio emission in jellyfish galaxies is still very much an open question.  As discussed above, adding more ancillary data for the jellyfish galaxies in this work (e.g.,\ optical IFU spectroscopy, high-resolution, multifrequency radio continuum imaging) will allow this question to be further explored in a resolved sense at high physical resolution.}

\subsection{\textcolor{black}{Summary}}\label{sec:summary}

\textcolor{black}{In this work we have presented the first identification of jellyfish galaxies in the nearby, massive Perseus cluster. These objects are identified on the basis of one-sided radio continuum tails, but also have rest-frame optical morphologies consistent with ongoing RPS.  Below we list the primary conclusions of this work.}
{\color{black}
\begin{enumerate}
    \item All four identified jellyfish galaxies show evidence of enhanced star formation on the leading side of the galaxy, opposite to the direction of the stripped tail. This is apparent from both visual inspection of the multiwavelength galaxy images and from a quantitative analysis of narrowband $\mathrm{H\alpha}$ sources (Figs~\ref{fig:panel_imgs}, \ref{fig:Ha_src_ang}). \\
    
    \item Between 144 MHz and 344 MHz, these Perseus jellyfish galaxies have flat spectral indices over the region of the galaxy disk that may be a result of ionization losses in a dense ISM. Across the stripped tails the spectral indices are not well constrained and within uncertainties are consistent with both the expected injection spectral index from star formation and with moderate steepening due to synchrotron aging (Figs~\ref{fig:lofar_vla_imgs}, \ref{fig:spec_index}). \\
    
    \item Globally, these galaxies have enhanced 144 MHz luminosities relative to the standard low-frequency radio luminosity versus SFR relation. A similar radio excess has been reported for other jellyfish galaxies, though the physical driver(s) of this effect is still an open question (Fig.~\ref{fig:L144_sfr}).
\end{enumerate}
\noindent
While this work covers the central $2^\circ \times 2^\circ$, there is still significant area between $\sim R_{500}$ and $\sim R_{200}$ in Perseus that is not covered by our current LOFAR 144 MHz image. This area will be imaged with LOFAR in the future, eventually covering the entire Perseus cluster and potentially identifying more jellyfish galaxies. This will permit a more complete census of RPS in Perseus and continue to build on our understanding of evironmentally driven galaxy evolution across the whole of the Perseus cluster.
}

\begin{acknowledgements}
We thank the anonymous referee for their constructive comments which have substantially improved this paper. IDR, RJvW, and RT acknowledge support from the ERC Starting Grant Cluster Web 804208.  AB acknowledges support from the VIDI research programme with project number 639.042.729, which is financed by the Netherlands Organisation for Scientific Research (NWO).  AI acknowledges the Italian PRIN-Miur 2017 (PI A. Cimatti).
\par
This paper is based on data obtained with the International LOFAR Telescope (ILT). LOFAR \citep{vanhaarlem2013} is the LOw Frequency ARray designed and constructed by ASTRON. It has observing, data processing, and data storage facilities in several countries, which are owned by various parties (each with their own funding sources) and are collectively operated by the ILT foundation under a joint scientific policy. The ILT resources have benefited from the following recent major funding sources: CNRS-INSU, Observatoire de Paris and Universit\'e d'Orl\'eans, France; BMBF, MIWF-NRW, MPG, Germany; Science Foundation Ireland (SFI), Department of Business, Enterprise and Innovation (DBEI), Ireland; NWO, The Netherlands; The Science and Technology Facilities Council, UK; Ministry of Science and Higher Education, Poland; The Istituto Nazionale di Astrofisica (INAF), Italy. This research made use of the Dutch national e-infrastructure with support of the SURF Cooperative (e-infra 180169) and the LOFAR e-infra group. The J\"ulich LOFAR Long Term Archive and the GermanLOFAR network are both coordinated and operated by the J\"ulich Supercomputing center (JSC), and computing resources on the supercomputer JUWELS at JSC were provided by the Gauss center for Supercomputinge.V. (grant CHTB00) through the John von Neumann Institute for Computing (NIC). This research made use of the University of Hertfordshire high-performance computing facility (\url{http://uhhpc. herts.ac.uk}) and the LOFAR-UK computing facility located at the University of Hertfordshire and supported by STFC [ST/P000096/1], and of the Italian LOFAR IT computing infrastructure supported and operated by INAF, and by the Physics Department of Turin University (under an agreement with Consorzio Interuniversitario per la Fisica Spaziale) at the C3S Supercomputing center, Italy.
\par
The Isaac Newton Telescope and its service mode are operated on the island of La Palma by the Isaac Newton Group of Telescopes in the Spanish Observatorio del Roque de los Muchachos of the Instituto de Astrofísica de Canarias.
\par
The National Radio Astronomy Observatory is a facility of the National Science Foundation operated under cooperative agreement by Associated Universities, Inc.
\par
This paper is based [in part] on data from the Hyper Suprime-Cam Legacy Archive (HSCLA), which is operated by the Subaru Telescope. The original data in HSCLA was collected at the Subaru Telescope and retrieved from the HSC data archive system, which is operated by the Subaru Telescope and Astronomy Data Center at National Astronomical Observatory of Japan. The Subaru Telescope is honored and grateful for the opportunity of observing the Universe from Maunakea, which has the cultural, historical and natural significance in Hawaii.
\par
This paper makes use of software developed for the Vera C. Rubin Observatory. We thank the observatory for making their code available as free software at  http://dm.lsst.org.
\par 
This publication makes use of data products from the Wide-field Infrared Survey Explorer, which is a joint project of the University of California, Los Angeles, and the Jet Propulsion Laboratory/California Institute of Technology, funded by the National Aeronautics and Space Administration.
\par
We acknowledge the usage of the HyperLeda database (http://leda.univ-lyon1.fr).
\end{acknowledgements}

%
%

\bibliographystyle{aa}
\bibliography{main}

\end{document}